\documentclass[12pt]{article}
\usepackage{epsfig}
\usepackage{graphicx}

\def\beq{\begin{equation}}
\def\eeq#1{\label{#1}\end{equation}}
\def\eeqn{\end{equation}}

\def\beqa{\begin{eqnarray}}
\def\eeqa#1{\label{#1}\end{eqnarray}}
\def\eeqan{\end{eqnarray}}

\let\bar=\overbar

\def\L{{\cal L}}

\def\Dslash{\not{\hbox{\kern-4pt $D$}}}
\def\dslash{\not{\hbox{\kern-2pt $\del$}}}

\def\msb{{\bar{\ssstyle M \kern -1pt S}}}

\usepackage{fancyhdr,graphicx}
\def\Title#1{\begin{center} {\Large {\bf #1} } \end{center}}

\begin{document}

\Title{Multiquark interactions and heavy hybrid stars}

\bigskip\bigskip


\begin{raggedright}

{\it 
Sanjin Beni\' c$^{1,2}$\\
\bigskip
$^{1}$Physics Department, Faculty of Science, 
University of Zagreb, Bijeni\v cka c.~32,
Zagreb 10000, Croatia\\
\bigskip
$^{2}$Department of Physics, The University of Tokyo, 7-3-1 Hongo, Bunkyo-ku, Tokyo 113-0033, Japan\\
}

\end{raggedright}

\section*{Abstract}

We introduce a two flavor Nambu--Jona-Lasinio (NJL) model with 8-quark interactions in the scalar and the vector channel.
With the lower density region described by the density-dependent relativistic mean field model we construct a hybrid equation of state.
We especially focus on the 4-quark vector couplings and the 8-quark vector NJL couplings and allocate a region in this parameter subspace where hybrid stars with masses larger than $2M_\odot$ exist.

\section{Introduction}

The recent discoveries of $2M_\odot$ compact stars \cite{Demorest:2010bx,Antoniadis:2013pzd} push 
us to understand the nature of matter at high densities.
It is possible that within 
very heavy compact stars there is a transition to quark matter.
We focus on hybrid stars, composed of a 
nuclear envelope and a quark core.

\begin{figure}[t]
\centering
\includegraphics[scale=0.35,clip]{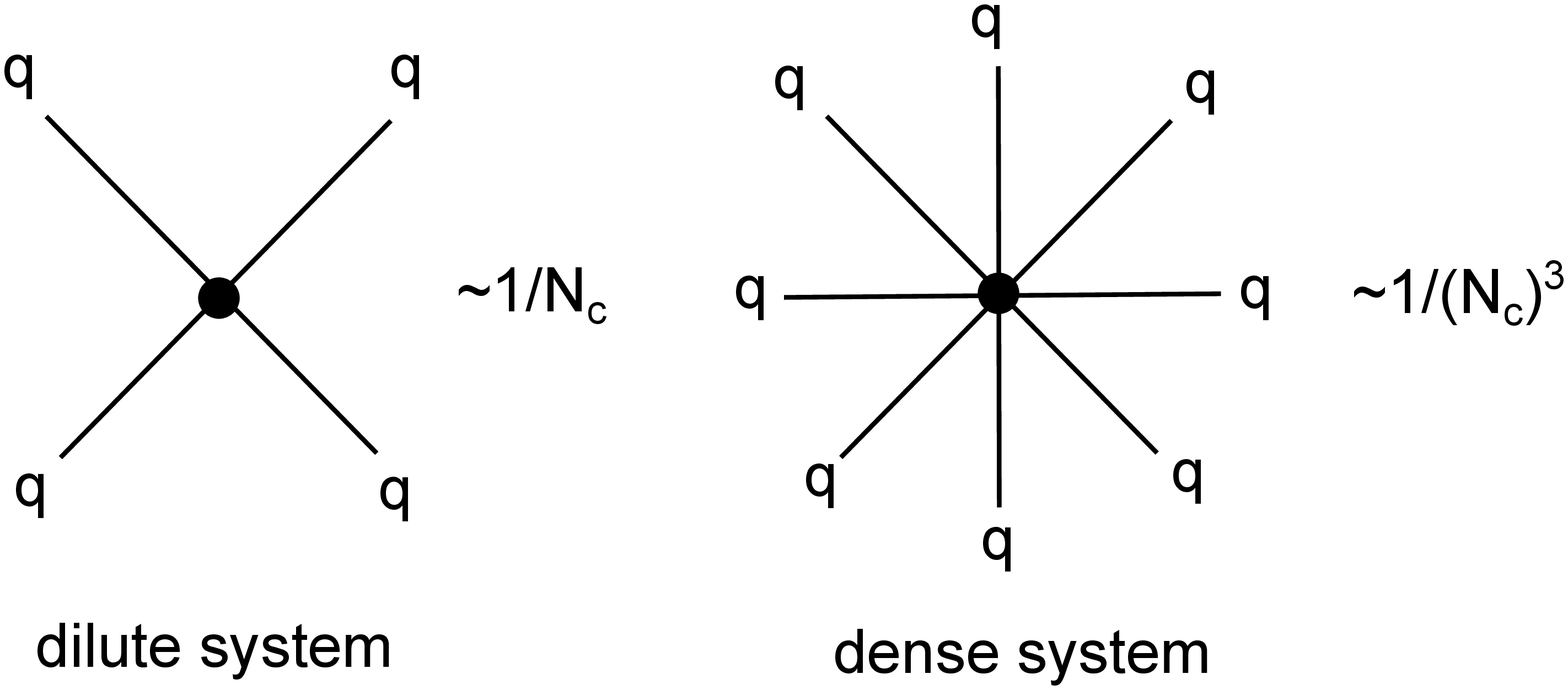}
\caption{$N_c$ scaling of 8-quark interactions.}
\label{fig:8q}
\end{figure}

In order to have a $2M_\odot$ star with 
a sizeable region of quark matter inside, quarks should have a 
low onset and the quark equation of state (EoS) 
should be stiff at high densities \cite{Alford:2013aca}.
There are several microscopic mechanisms to 
introduce such a situation \cite{Klahn:2006iw,Pagliara:2007ph,Weissenborn:2011qu,Lenzi:2012xz,Klahn:2013kga}.
In this work we introduce one such mechanism: 8-quark interactions within the 
NJL framework \cite{Benic:2014iaa,Benic:2014jia}.
Although the 8-quark coupling naively scales 
as $1/N_c^2$ with respect to the 4-quark coupling (see Fig.~\ref{fig:8q}), in a 
dense system the 
expectation value of
the corresponding operator (e.~g. the vector current operator) 
can be large. 
Therefore, 
8-quark interactions should become 
important at high densities.

The hybrid equation of state (EoS) model has several parameters; 
we have considered a subspace of vector
interactions in the quark sector.
These are the 
4-quark and the 8-quark vector 
interaction, quantified by 
dimensionless couplings, $\eta_2$ and $\eta_4$, respectively.
There are in general large uncertainties concerning 
the vector interactions for quark matter.
We find a parameter space that allows 
for hybrid stars with masses higher than $2M_\odot$.

\section{Hybrid equation of state model}

We consider a two flavor model of matter in beta equilibrium.
Nuclear matter is described with the DD2 EoS, for 
details see \cite{Typel:2009sy}.
Quark matter EoS is taken to be the 
NJL model with 8-quark interactions (NJL8)
in the vector and in the 
scalar channel.
The thermodynamic potential of the NJL8 model is
\begin{equation}
\Omega = U+ \sum_{f=u,d}\Omega_f(M_f,T,\tilde{\mu}_f)-\Omega_0~,
\end{equation}
where the classical and the one-loop contribution are
\begin{equation}
U = 2\frac{g_{20}}{\Lambda^2}(\phi_u^2+\phi_d^2) + 
12 \frac{g_{40}}{\Lambda^8}(\phi_u^2+\phi_d^2)^2
-2\frac{\eta_2 g_{20}}{\Lambda^2}(\omega_u^2 + \omega_d^2)-
12 \frac{\eta_4 g_{40}}{\Lambda^8}(\omega_u^2+\omega_d^2)^2~,
\end{equation}
\begin{equation}
\Omega_f = -2 N_c\int \frac{d^3 p}{(2\pi)^3}
\left\{E_f + T\log[1+e^{-\beta(E_f-\tilde{\mu}_f)}]+ T\log[1+e^{-\beta(E_f+\tilde{\mu}_f)}]\right\}~,
\end{equation}
and $\phi_f$ and $\omega_f$ are scalar and vector mean-fields.
More details on the model can be 
found in Refs.~\cite{Benic:2014iaa,Benic:2014jia}
and references therein.
The nuclear-quark phase transition is assumed to be first order.
The important point about the nuclear sector is
that DD2 is itself quite stiff, so maximum mass of a 
pure neutron star is already well above $2M_\odot$.
In the quark sector 
the key role is played by the 8-quark vector interactions.
Since they become relevant 
only at higher densities, the stiffness of quark
matter is pronounced at higher densities, see Fig.~\ref{fig:mr}.
On the other hand, the onset of quark matter is controlled by the 4-quark vector interactions.

\section{Masses and radii}

\begin{figure}[t]
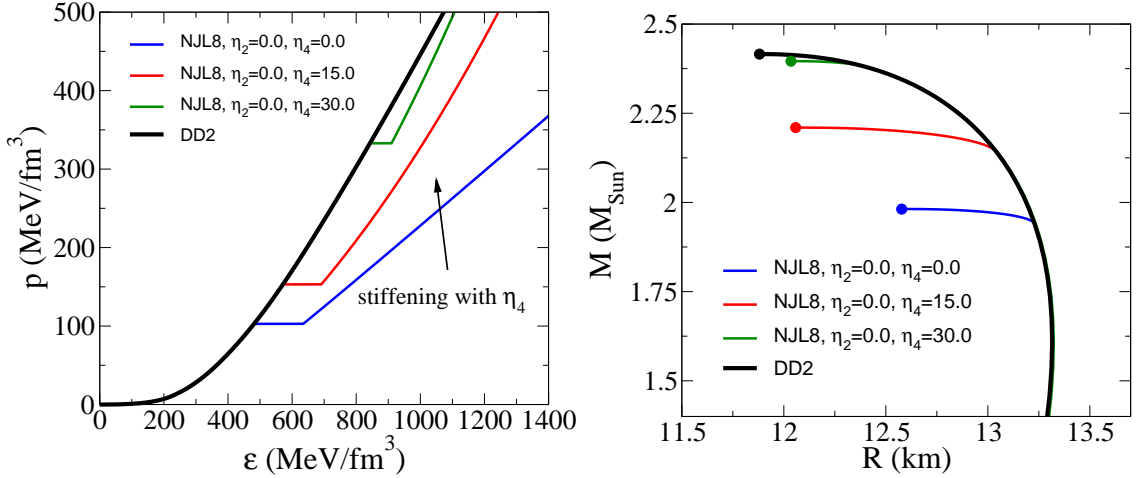

\centering
\includegraphics[scale=0.45,clip]{eos_hyb2.eps}
\vspace{0.3cm}
\includegraphics[scale=0.45,clip]{tov_hyb.eps}
\caption{Left panel: EoS, 
Right panel: $M-R$ relationship.
Results are shown for fixed $\eta_2=0.0$ and several $\eta_4$.}
\label{fig:mr}
\end{figure}
We find that without vector interactions it is not 
possible to consider a hybrid star with $2 M_\odot$ within this model.
With finite vector interactions, several 
options for achieving $M>2M_\odot$ become viable.
For example, on Fig.~\ref{fig:mr} we show the effect 
of 8-quark vector 
interaction on the maximum mass of a hybrid star.
On Fig.~\ref{fig:e24} the possible star configurations 
within a parametric region spanned by $\eta_2$ and $\eta_4$ are shown.
The lower values of $\eta_2$ and $\eta_4$ are excluded by the $2M_\odot$ constraint.
Then, the upper bounds will in general be governed 
by causality, instability caused by the softening of the EoS at the transition, or the masquerade effect \cite{Alford:2004pf}.
In the direction $\eta_2=0$ and $\eta_4\geq 0$ we find the masquerade effect, which is also visible on Fig.~\ref{fig:mr}.
In the opposite direction $\eta_4=0$ and $\eta_2\geq 0$
the appearance of quark matter makes the star unstable.
However, within a limited window of $\eta_2$ it is still 
possible to recover a stable hybrid configuration 
by increasing $\eta_4$.
The speed of sound at densities reached in the cores of 
all stable configurations is below unity.

Note that in \cite{Benic:2014jia} the DD2 EoS was adjusted by an excluded volume contribution, but this is not the approach considered here.
The main result found there was a large increase in the latent heat leading to third family hybrid stars in the mass-radii diagram.

\section{Conclusion}

\begin{figure}[t]
\centering
\includegraphics[scale=0.6,clip]{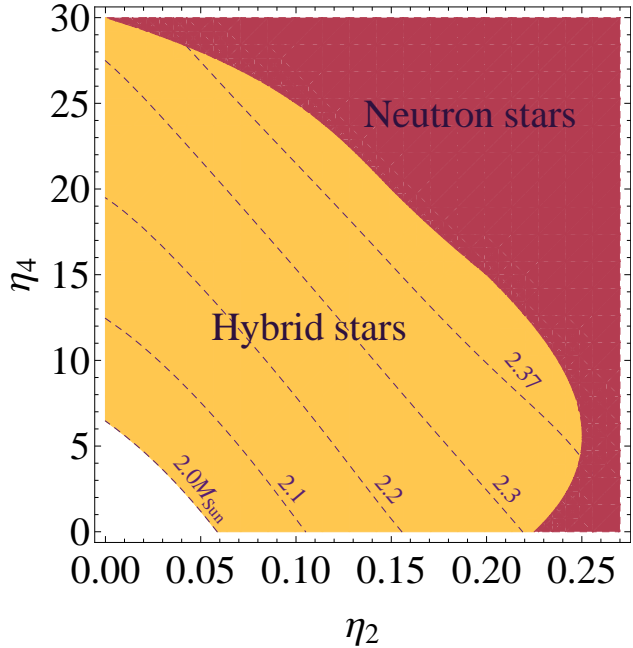}
\caption{Composition of stars in the $\eta_2-\eta_4$ plane.
The curves denote maximum masses for hybrid stars.}
\label{fig:e24}
\end{figure}
We have shown that heavy hybrid stars can 
be obtained within the NJL model with 8-quark interactions.
They represent a microscopic way of modeling very 
stiff quark matter at densities realizable in compact stars.
The parametric diagram considered 
on Fig.~\ref{fig:e24} is very sensitive to the details
of the nuclear-quark transition.
For example, increasing the latent heat of the transition
brings an interesting possibility of 
twin stars at $2M_\odot$ \cite{Benic:2014jia}.  
It will be interesting 
to explore the role of strangeness \cite{Moreira:2014qna}.
Finally, the 
parameter window obtained here is likely to be reduced by 
further constraints on 
the vector interactions in quark matter \cite{Steinheimer:2014kka,Alvarez-Castillo:2014nua}.

\subsection*{Acknowledgement}

I would like to thank the 
organizers of the CSQCD IV conference for providing an 
excellent atmosphere.
This work comes from many discussions with D.~Blaschke and D.~E.~A.~Castillo, for which I am grateful. 
I have also benefited from discussions with T.~Fischer and S.~Typel.
S.~B. is partially supported by the Croatian Science Foundation under Project No. 8799 and by the COST Action MP1304 NewCompStar.

\end{document}